# Bounds imposed on the sheath velocity of a dense plasma focus by conservation laws and ionization stability condition

S K H Auluck[1]

Physics Group, Bhabha Atomic Research Center, Mumbai, INDIA

Abstract:

Experimental data compiled over five decades of dense plasma focus research is consistent with the snowplow model of sheath propagation, based on the hypothetical balance between magnetic pressure driving the plasma into neutral gas ahead and "wind pressure" resisting its motion. The resulting sheath velocity, or the numerically proportional "drive parameter", is known to be approximately constant for devices optimized for neutron production over 8 decades of capacitor bank energy. This paper shows that the validity of the snowplow hypothesis, with some correction, as well as the non-dependence of sheath velocity on device parameters, have their roots in local conservation laws for mass, momentum and energy coupled with the ionization stability condition. Both upper and lower bounds on sheath velocity are shown to be related to material constants of the working gas and independent of the device geometry and capacitor bank impedance.

The dense plasma focus[1]  (DPF) is a laboratory plasma fusion device with a rich and complex phenomenology [2,3]; yet it also has strikingly simple universal properties which hold over 8 decades in capacitor bank storage energy: neutron yield Y scales as the fourth power of pinch current $I_p$ [3,4], the ratio of capacitor bank energy to the cube of anode radius is nearly constant [4] and the drive parameter $I / p_0 \sqrt{a}$ , with $p_0$ the fill gas pressure, $I$ the maximum current and $a$ the anode radius, is nearly constant [4,5]. One of the conclusions of DPF research over the last 50 years is that the plasma current sheath (PCS) propagation is quite well described by the snowplow model [6,7,8], which assumes that the sheath acquires a velocity and shape dictated by a balance between magnetic pressure $B^2/2\mu_0$ driving the plasma into the neutral gas and "wind pressure" $\rho_0 v^2$ resisting it: $v = B/\sqrt{2\mu_0\rho_0}$ . The well-known Lee model [9] uses the snowplow hypothesis to fit experimental current waveforms during the rundown phase by a judicious choice of a few adjustable parameters. After incorporation of circuit resistance [10], the Gratton-Vargas (GV) two dimensional analytical snowplow model [11] is also able to reproduce the experimental current waveforms in a number of facilities, again by choosing numerical values of the gas pressure, static inductance and circuit resistance as fitting parameters. The GV model [11] has also been used to calculate [12] the electromagnetic work done per unit swept mass $\varepsilon_{EM}$ in certain experiments. This quantity is shown [12] to decrease below the specific ionization energy $\varepsilon_i$ for hydrogen with increasing operating pressure and this transition is found to be correlated with a relatively rapid decrease of neutron yield above a certain pressure. An extension of this study [13] suggests that occurrence of compression ratio much higher than the value 4 for an ideal adiabatic shock is related with the increase of $\varepsilon_{EM}$ over $\varepsilon_i$.

[1] Postal address: Physics Group, Bhabha Atomic Research Center, Mumbai-400085, INDIA:
E-mail: skhauluck@gmail.com; skauluck@barc.gov.in

The sheath velocity at the anode surface predicted by the snowplow model happens to be proportional to the drive parameter. One of the open questions [14] of DPF research is: why does the DPF operate in a narrow range of velocities? A number of possible answers are offered [5] but as yet there is no theoretical analysis of the bounds on the velocity of the DPF plasma current sheath. Klir and Soto [14] have used non-dimensional rendering of the Lee model [9] to argue for the existence of a similarity relationship concerning evolution of the axial phase between devices having equal values of a dimensionless parameter involving device geometry, operating pressure and capacitor bank impedance and its relation with the upper limit of sheath velocity in neutron optimized devices. They suggest that this upper limit may be a consequence of the DPF optimizing procedure rather than a fundamental property of the plasma and as such, could be overcome in a different but related device, such as a gas-puff pinch, which has a larger number of independent control parameters.

This Letter demonstrates that the origin of the snowplow effect and the narrow range of sheath velocity lies in conservation laws for mass, momentum and energy, including the contribution from the electromagnetic field [15,16,17], applied to the PCS configuration consisting of an azimuthal magnetic field and plasma in the downstream zone and a neutral gas without magnetic field in the upstream zone, and on an ionization stability condition already known in the theory of ionizing MHD shocks [16]. In contrast with arguments [5,12,13,14] based on integrated properties of the DPF in the snowplow model, which include geometrical aspects of the plasma device and electrical properties of the capacitor bank, the present approach uses local properties of the plasma. As a result, both the upper and lower bounds on velocity are shown to be related with material constants of the working gas and to be independent of the capacitor bank impedance and DPF geometry.

The manifestly curved shape [3,18] of the DPF plasma usually poses a major mathematical problem in using the conservation law formalism. However, the assumption of effectively plane geometry in generating the jump conditions across a shock discontinuity in such analysis can be justified for the curved geometry of the PCS by arguments similar to those of Libermann and Velikhovich [16] and Goedbloed [17] namely that the curvature can be neglected over transverse dimensions larger than the thickness of the shock discontinuity decided by dissipative phenomena but much less than the radius of curvature. This simplification should hold for DPF operation for times sufficiently before the pinch phase.

The conservation of mass, momentum and energy, including contributions from the electromagnetic field, is described by the following equations 1, in which $\rho$ is the mass density, $\vec{v}$ is the fluid velocity, $p$ is the fluid pressure, $\vec{B}, \vec{E}$ are the magnetic and electric fields, $\mathcal{E}$ is the total energy density, $\mathcal{U}_L'$ is the loss of energy per unit volume per unit time through processes such as radiation and thermal conduction and $\overline{\overline{\mathbb{I}}}$ is a unit dyadic:

$$\partial_t \rho + \vec{\nabla} \cdot (\rho \vec{v}) = 0$$
$$\partial_t \left( \rho \vec{v} + \vec{E} \times \vec{B}/\mu_0 c^2 \right) + \vec{\nabla} \cdot \left( \rho \vec{v}\vec{v} + \overline{\overline{\mathbb{I}}} \left( p + B^2/2\mu_0 \right) - \vec{B}\vec{B}/\mu_0 \right) = 0 \qquad 1$$
$$\partial_t \mathcal{E} + \vec{\nabla} \cdot \left( \mathcal{E}\vec{v} + p\vec{v} + \vec{E} \times \vec{B}/\mu_0 \right) = -\mathcal{U}_L'$$

Note that resistive dissipation of energy is already taken into account in 1; this comes about from the conservation law form of electromagnetism which expresses $\vec{J}.\vec{E}$ in terms of the rate of change of electromagnetic energy density and divergence of the Poynting vector. The relation between the pressure and density must take into account the fact that the gas has variable degrees of dissociation $\delta$ and ionization $\alpha$ across the shock front, varying from 0 in upstream region to nearly 1 in the downstream region. The upstream region has neutral diatomic gas while the downstream region has fully dissociated and almost fully ionized mono-atomic plasma with respective pressures given by:

$$p_0 = \rho_0 T_0 / 2m_i \; ; \; p_1 = \rho_1 T_1 \left(1 + \alpha_1\right) / m_i \qquad\qquad 2$$

This makes the simplifying assumption that electrons, ions (of mass $m_i$) and diatomic neutrals (of mass $2m_i$) all have the same temperature T in energy units; essentially this assumes that a local thermal equilibrium is established over the time scale of sheath propagation (0.1-1 μs). The total energy $\mathcal{E}$ per unit volume includes the internal energy density (the sum of ionization and dissociation energy density and the internal thermal energy density), the kinetic energy density and the magnetic energy density (electric energy density is negligible in quasi-neutral plasma and neutral gas):

$$\mathcal{E} = \left(\alpha\rho\varepsilon_i + \delta\rho\varepsilon_d + p / \left(\gamma - 1\right)\right) + \rho v^2 / 2 + B^2 / 2\mu_0 \qquad\qquad 3$$

Its flux involves convection of the total energy $\mathcal{E}\vec{v}$, work done in compression $p\vec{v}$ and the Poynting vector. Since the downstream state is (almost) fully ionized hot plasma, the electric field is given by $\vec{E} = -\vec{v} \times \vec{B}$, determining the Poynting vector for the assumed configuration of the PCS as $\vec{E} \times \vec{B} / \mu_0 = \vec{v}\, B^2 / \mu_0 - \vec{B} \left(\vec{v} \cdot \vec{B}\right) / \mu_0$. For shock and material velocities much less than the velocity of light, the electromagnetic momentum density can be neglected in the second of equations 1. In 3, $\varepsilon_d$ is the energy per unit mass required to fully dissociate the (usually diatomic) neutral gas and $\varepsilon_i$ is the energy per unit mass required to convert it into fully ionized mono-atomic plasma. Taking the molecular dissociation energy of hydrogen as 4.5 eV and atomic ionization energy as 13.6 eV, $\varepsilon_d \approx 1.05 \times 10^8$ J/kg, $\varepsilon_i \approx 6.4 \times 10^8$ J/kg for deuterium gas. The polytropic index is $\gamma = 7/5$ for the diatomic gas upstream while it is $\gamma = 5/3$ for the mono-atomic gas downstream.

The inhomogeneous term $\mathscr{U}_L^l$ is usually treated [19] by the methods of radiation hydrodynamics. The present discussion is limited to the homogeneous problem.

The analysis is facilitated by constructing a local curvilinear coordinate system $(\zeta, \theta, \xi)$ in terms of a unit vector $\hat{\zeta}$ along the tangent to sheath, a unit vector $\hat{\theta}$ along the azimuth and a unit vector perpendicular to the sheath defined as $\hat{\xi} \equiv \hat{\zeta} \times \hat{\theta}$. The locally unidirectional nature of the flow (along the $\hat{\xi}$ direction) implies that local variations of the quantities in the two transverse directions $\hat{\zeta}$ and $\hat{\theta}$ are negligible in the first order, which could be determined, if needed, by successive approximation in a subsequent iteration. The divergence operator can then be replaced by a single partial space derivative in the direction of flow, neglecting the curvature over sufficiently short distance scales in the transverse direction which are nevertheless larger than the thickness of the dissipative layer [16]. The partial differential equations 1 then form a system of N (=3) hyperbolic conservation laws of the form $\partial_t u_i + \partial_\xi f_i = w, i = 1 \cdots N$ having essential nonlinearity [19] which are known to possess discontinuous travelling wave solutions: functions of $\xi - st$, where $s$ is the speed of the travelling discontinuity and which satisfy the Rankine-Hugoniot (RH) conditions $s[u_i] = [f_i]$ at every discontinuity. The square bracket here represents the difference of the enclosed quantity across the discontinuity; upstream and downstream states are labeled by subscripts 0 and 1. For the homogeneous part of equations 1 [19], the RH conditions become

$$
\begin{aligned}
& s\left(\rho_1 - \rho_0\right) = \left(\rho_1 v_1 - \rho_0 v_0\right) \\
& s\left(\rho_1 v_1 - \rho_0 v_0\right) = \rho_1 v_1^2 - \rho_0 v_0^2 + p_1 - p_0 + B_1^2/2\mu_0 \\
& s\left(\rho_1 \alpha_1 \varepsilon_i + \rho_1 \varepsilon_d + \left(p_1 - p_0\right)/(\gamma - 1) + \left(\rho_1 v_1^2 - \rho_0 v_0^2\right)\big/2 + B_1^2/2\mu_0\right) \\
& = \rho_1 \alpha_1 \varepsilon_i v_1 + \rho_1 \varepsilon_d v_1 + \left(p_1 v_1/(\gamma_1 - 1) - p_0 v_0/(\gamma_0 - 1)\right) + \left(\rho_1 v_1^3 - \rho_0 v_0^3\right)\big/2 + \left(p_1 v_1 - p_0 v_0\right) \\
& + v_1 B_1^2/2\mu_0 + v_1 B_1^2/\mu_0
\end{aligned}
\qquad 4
$$

In 4, the last term on the right hand side of the third equation is the Poynting vector; the term before that is the convection of magnetic energy density by the fluid. The absence of magnetic field in the upstream state is explicitly taken into account. The downstream degree of ionization $\alpha_1$ may be slightly different from 1.

Defining specific volume $V \equiv \rho^{-1}$, $\varepsilon_{eff} \equiv \alpha_1 \varepsilon_i + \varepsilon_d$, and assuming upstream fluid to be at rest, the following results, analogous to the theory of shock waves in condensed energetic materials, can be obtained by standard algebraic manipulations,:

The Rayleigh line: $\dfrac{p_1 + B_1^2/2\mu_0 - p_0}{V_1 - V_0} = -s^2 \rho_0^2$          5

The Hugoniot:

$$\left( \frac{\gamma_1}{\gamma_1 - 1} p_1 V_1 - \frac{\gamma_0}{\gamma_0 - 1} p_0 V_0 \right) - \frac{1}{2}\left( p_1 - p_0 \right)\left( V_1 + V_0 \right) + \frac{B_1^2}{2\mu_0}\left( \frac{5}{2} V_1 - \frac{3}{2} V_0 \right) + \varepsilon_{eff} = 0 \qquad 6$$

These equations reduce to the conventional equations for a gas dynamic shock when $B_1^2/2\mu_0$ and $\varepsilon_{eff}$ are equated to zero and for a gas dynamic detonation when $B_1^2/2\mu_0$ is zero and $\varepsilon_{eff}$ is negative for an exothermic reaction. They are rendered dimensionless by defining

$$\tilde{p} \equiv \frac{p}{B_1^2/2\mu_0}; \tilde{V} \equiv \frac{V}{V_0}; \tilde{s}^2 \equiv \frac{s^2}{v_{sp}^2}; \tilde{\varepsilon}_{eff} \equiv \frac{\varepsilon_{eff}}{v_{sp}^2} \qquad 7$$

The "snowplow velocity" is defined as $v_{sp} \equiv B_1/\sqrt{2\mu_0\rho_0}$. A switch $\Sigma$ (equal to 1 or 0) is left as a placeholder for $B_1^2/2\mu_0$; this is used to compare the case of a non-zero downstream magnetic field with the case of a usual gas dynamic shock, in which case the normalization of pressure by $B_1^2/2\mu_0$ just amounts to a change of units. Note that $\tilde{\varepsilon}_{eff}$ varies as $r^2$ across the PCS because of the $r^{-1}$ variation of the magnetic field with radius $r$. The dimensionless working relations are then

$$\frac{\tilde{p}_1 + \Sigma - \tilde{p}_0}{1 - \tilde{V}_1} = \tilde{s}^2 \qquad 8$$

$$\left(\frac{\gamma_1}{\gamma_1-1}\,\tilde{p}_1\tilde{V}_1 - \frac{\gamma_0}{\gamma_0-1}\,\tilde{p}_0\right) - \frac{1}{2}\left(\tilde{p}_1 - \tilde{p}_0\right)\left(\tilde{V}_1 + 1\right) + \Sigma\left(\frac{5}{2}\tilde{V}_1 - \frac{3}{2}\right) + \tilde{\varepsilon}_{eff} = 0 \qquad\qquad 9$$

The explicit $\left(\tilde{p}, \tilde{V}\right)$ form of the Hugoniot is

$$\tilde{p}_1 = \frac{\tilde{p}_0\tilde{V}_1 - \Gamma_0\tilde{p}_0 + \Sigma\left(5\tilde{V}_1 - 3\right) + 2\tilde{\varepsilon}_{eff}}{\left(1 - \Gamma_1\tilde{V}_1\right)}; \Gamma \equiv \left(\gamma + 1\right)/\left(\gamma - 1\right) \qquad\qquad 10$$

For the gas dynamic, non-ionizing shock, $\Sigma = 0, \tilde{\varepsilon}_{eff} = 0$, the well-known divergence of the Hugoniot at a compression ratio $\sim \Gamma_1$ is observed, leading to a maximum compression ratio of 4 for an ideal adiabatic shock [13] in a mono-atomic gas. This persists with the case of $\Sigma = 1, \tilde{\varepsilon}_{eff} \neq 0$ leading to an infinite value of $\tilde{p}_1$ from 10, and therefore of $\tilde{s}^2$ from 8, for $\tilde{V}_1 = \Gamma_1^{-1}$. Curiously, a solution with both $\tilde{V}_1$ and $\tilde{p}_1$ equal to zero is allowed by the conservation laws for $\Sigma = 1, \tilde{\varepsilon}_{eff} > 0$: for this case, 8 reduces to $\tilde{s} = \sqrt{1 - \tilde{p}_0}$, 10 reduces to $\tilde{\varepsilon}_{eff} = 3/2 + \tilde{p}_0\Gamma_0/2$. This is the limiting case of an ideal ionizing snowplow (IIS), where energy dissipated in the shock is barely sufficient to dissociate and ionize the medium, sparing no energy for heating so that $\tilde{p}_1 \approx 0$. At the same time, the ionized material sticks to the shock front, having no spare kinetic energy to move away from it, resulting in unrestricted pile-up leading to $\tilde{V}_1 \approx 0$. Such ideal ionizing snowplow is disallowed by the second law of thermodynamics which requires the entropy change $\Delta\mathbb{S} = c_v \ln\left(p_1 V_1^{\gamma} / p_0 V_0^{\gamma}\right) \geq 0$. It can be easily shown from 8 that entropy-increasing shock wave solutions consistent with conservation laws for $\tilde{p}_1 - \tilde{p}_0 > 0$ and $\tilde{V}_1 > 0$, *must* travel with $\tilde{s} \geq 1$. The quantity $\Gamma_1 \equiv \left(\gamma_1 + 1\right)/\left(\gamma_1 - 1\right)$ represents the *maximum compression* $(\tilde{V}_1^{-1})$ in a single gas dynamic shock travelling at $\tilde{s} \to \infty$; in contrast a snowplow shock travelling at $\tilde{s} = 1$ would have infinite compression.

In the practical case of DPF, the upstream pressure $\sim$ few mbar and the magnetic field $\sim 10$ T; as a result $\tilde{p}_0 \sim 10^{-5}$ which may be neglected in comparison with other terms. Elimination of $\tilde{p}_1$ between 8 and 9 (for $\Sigma = 1$) leads to a quadratic equation for $\tilde{V}_1$ in terms of $\tilde{s}^2$ and $\tilde{\varepsilon}_{eff}$ so that two solutions are obtained for $\tilde{V}_1$ and $\tilde{p}_1$.

$$\tilde{V}_1 = \frac{1 + 5\tilde{s}^2 \mp \sqrt{9\tilde{s}^4 + \tilde{s}^2\left(-22 + 32\tilde{\varepsilon}_{eff}\right) + 1}}{8\tilde{s}^2}$$

$$\tilde{p}_1 = \frac{1}{8}\left(-9 + 3\tilde{s}^2 \pm \sqrt{9\tilde{s}^4 + \tilde{s}^2\left(-22 + 32\tilde{\varepsilon}_{eff}\right) + 1}\right)$$

$\qquad\qquad 11$

The bottom signs of the square roots must be rejected since they lead to negative pressure for $\tilde{\varepsilon}_{eff} > 0$. For case of $\tilde{\varepsilon}_{eff} < 0$ (exothermic reaction), the bottom signs lead to the well-known deflagration solution. The downstream temperature $T_1$ obtained from 2 and 11 is

$$T_1 = \frac{1}{\left(1 + \alpha_1\right)} m_i p_1 V_1 \approx \frac{1}{2} m_i v_{sp}^2 \tilde{p}_1 \tilde{V}_1 \qquad\qquad 12$$

In order for the downstream plasma to have a degree of ionization $\alpha \approx 1$, its temperature $T_\alpha$ as given by the Saha ionization equation should be sufficiently high:

$$\frac{\alpha^2}{1 - \alpha} = \frac{m_i}{\rho_0}\left(\frac{2\pi m_e T_\alpha}{h^2}\right)^{3/2} \exp\left(-\varepsilon_H / T_\alpha\right) \quad \varepsilon_H = 13.6 \text{ eV} = 1 \text{ Ryd} \qquad 13$$

Defining $\tilde{T}_\alpha \equiv T_\alpha / \varepsilon_H$ , $\rho_H \equiv m_i\left(2\pi m_e \varepsilon_H / h^2\right)^{3/2} \sim 500 \text{ kg/m}^3$, $\bar{\alpha} \equiv 1 - \alpha \ll 1$ the Saha equation can be written in dimensionless form as

$$f\left(\tilde{T}_\alpha\right) \equiv \tilde{T}_\alpha^{-3/2} \exp\left(1/\tilde{T}_\alpha\right) = \frac{\bar{\alpha}}{\rho_0 / \rho_H} \equiv \Lambda \qquad\qquad 14$$

Clearly, the downstream temperature $T_1$ given by 12 should exceed the temperature $T_{\alpha 1}$ necessary to maintain the degree of ionization $\alpha_1$

$$\tilde{p}_1 \tilde{V}_1 \geq \tilde{T}_{\alpha_1} \frac{\left(1 + \alpha_1\right)\tilde{\varepsilon}_{eff}}{\alpha_1 + m_i \varepsilon_d / \varepsilon_H} \qquad\qquad 15$$

Fig 1 shows that for a wide range of parameter $\Lambda$, it is sufficient to assume $\tilde{T}_\alpha \geq 1$ to have a fully ionized downstream plasma

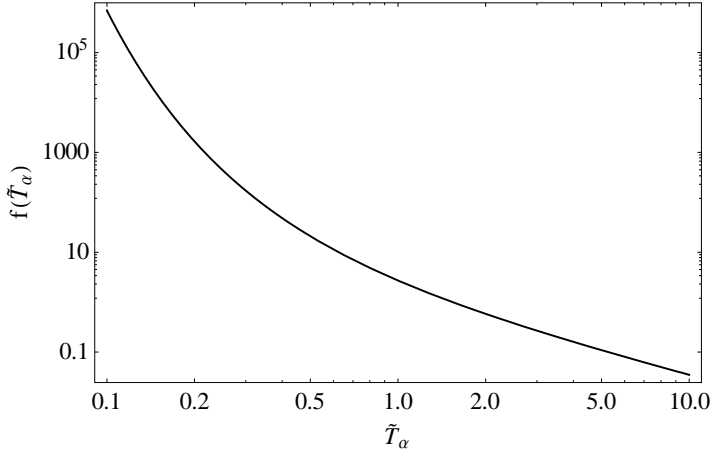

Fig 1: Saha equation 14 plotted in dimensionless form.

From 15, this gives

$$\tilde{p}_1 \tilde{V}_1 \geq 1.717 \tilde{\varepsilon}_{eff} \qquad\qquad 16$$

It is well-known [16] that conservation laws and equation of state do not completely define the downstream state of an ionizing shock wave. Theory of ionizing MHD shocks [15,16,17] reveals that the upstream transverse electric field plays an important role in determining the downstream state via the ionizing stability condition. Electromagnetic boundary conditions stipulate equality of the upstream and downstream transverse electric fields; the latter is given by $-v_1 B_1$ and the downstream velocity is determined by the equation of mass conservation to be $v_1 = s\left(1 - \tilde{V}_1\right)$. The ionization stability condition [16] stipulates that the upstream transverse electric field $\left|E_0\right| = \tilde{s}\left(1 - \tilde{V}_1\right)\sqrt{2\mu_0 \rho_0} v_{sp}^2$ must be less than or equal to a threshold breakdown electric field $E_b$; otherwise an ionization wave runs *ahead* of the shock wave [16] and not *with* it.

A simple estimate of $E_b$ is obtained as follows. The probability that a photoelectron generated in the upstream gas of molecular number density $n_0$ causes impact ionization while moving a distance $dx$ in the electric field is given by $dP_{im} = \sigma_{im}(w) n_0 dx$, where $\sigma_{im}(w)$ is the electron impact ionization cross section [20] as a function of electron energy $w$ (in eV), which is related to the electric field by $dw = E_b dx - w n_0 \sigma_c \, dx$, $\sigma_c$ being the collision cross-section in the neutral gas. Electron multiplication is guaranteed when the integrated probability equals unity; this condition can be written as

$$\int dP_{im} = \int \frac{\sigma_{im}(w)\,dw}{\left(E_b/n_0 - w\sigma_c(w)\right)} = 1 \qquad\qquad 17$$

This shows that $E_b/n_0$ is a material property of the gas. Using electron impact ionization cross-section data from the NIST database [20] and total electron collision cross-section in hydrogen from Yoon et. al. [21], this is found to be $\kappa \equiv E_b/n_0 \approx 8.21 \times 10^{-18} \cdot$ Volt-m$^2$ for hydrogen.

Introducing the ratio $\Theta \equiv E_b/\sqrt{2\mu_0\rho_0}\,v_{sp}^2 = \kappa n_0 \sqrt{2\mu_0\rho_0}/B_1^2$, the ionization stability condition $E_0 \leq E_b$ can be written as

$$\tilde{s}\left(1 - \tilde{V}_1\right) \leq \Theta \qquad\qquad 18$$

Fig. 2 shows the region in $\left(\tilde{\varepsilon}_{eff}, \tilde{s}\right)$ parameter space in which conditions 16 and 18 are both satisfied. It is found that there are no solutions for $\Theta < 0.872, \tilde{T}_\alpha \geq 1$. This implies a bound

$$\frac{\kappa\sqrt{\rho_0}}{2m_i\sqrt{2\mu_0}} > 0.872 v_{sp}^2 \ \text{ or } \ 1.08 \times 10^5 \cdot \sqrt[4]{p_0\,(mbar)} > v_{sp} \ \text{or} \ 0.064 \Big/ \sqrt{p_0\,(mbar)} < \tilde{\varepsilon}_{eff} \qquad 19$$

Fig. 2 shows an upper bound $\tilde{\varepsilon}_{eff} \leq 0.12$ for $\Theta = 0.872$, which translates to a lower bound on the snowplow speed: $v_{sp} > \sqrt{\varepsilon_{eff}/0.12} \sim 7.9 \times 10^4$ m/s. This agrees with the estimate [14] $7 \times 10^4$ m/s from the average value of driver parameter in a compilation of DPF data from many facilities. The upper bound 19 on $v_{sp}$ coincides with this lower bound at a fill pressure of 0.286 mbar of deuterium; this would then represent the lower limit to the operating pressure of a deuterium plasma focus. An accurate determination of this limit, along with this discussion, may facilitate experimental estimate of the lower limit on the downstream temperature.

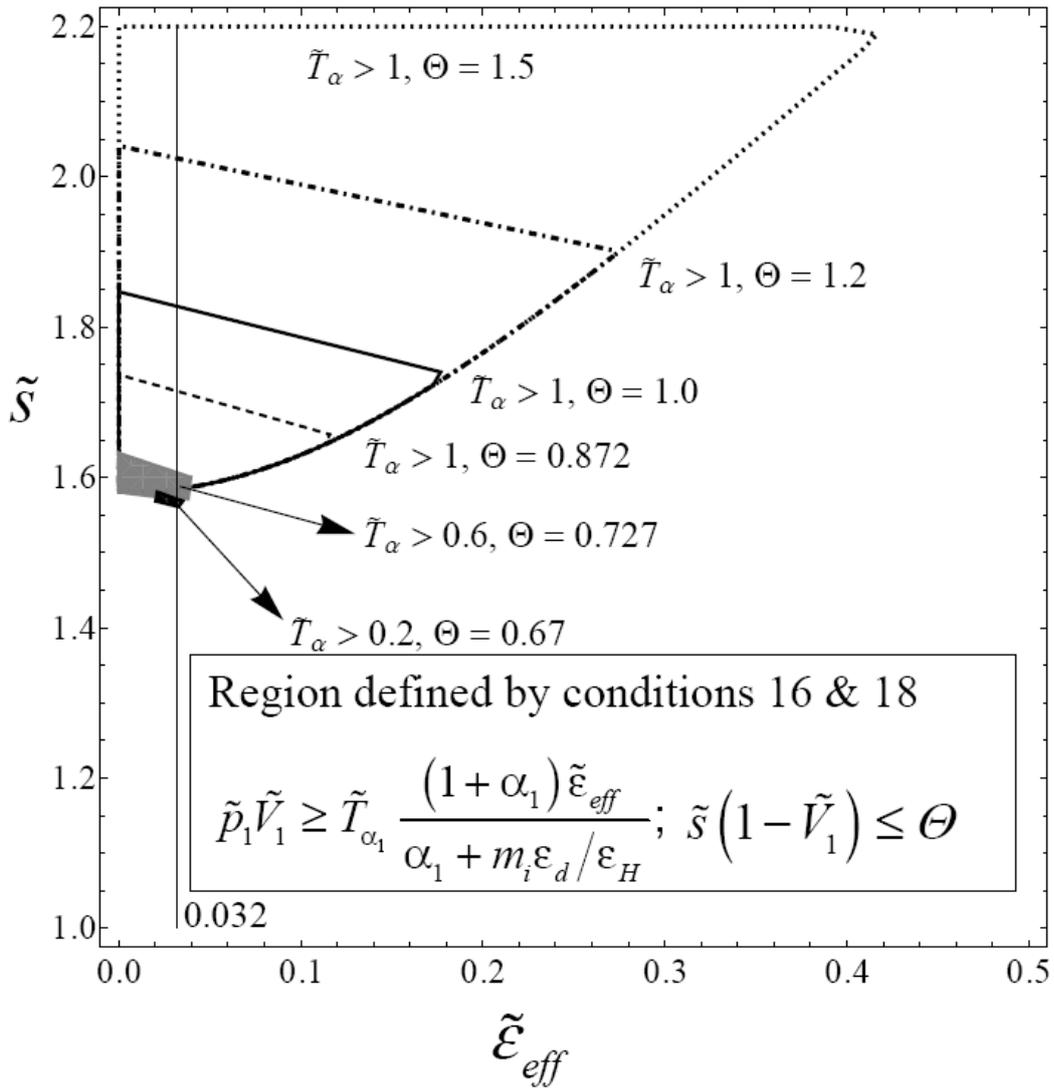

Fig.2: Region in $\left(\tilde{\varepsilon}_{eff}, \tilde{s}\right)$ parameter space in which conditions 16 and 18 are simultaneously satisfied. Four cases are shown in dashed $\left(\Theta = 0.872\right)$, solid $\left(\Theta = 1\right)$, dot-dashed $\left(\Theta = 1.2\right)$ and dotted $\left(\Theta = 1.5\right)$ lines where the downstream temperature is assumed to be greater than 1 Rydberg $\left(\tilde{T}_{\alpha} \geq 1\right)$. There are no solutions which satisfy both these conditions for $\Theta < 0.872$.

Two cases are shown for comparison in grey and black shaded regions where the downstream temperature is assumed to be greater than 0.6 Rydberg ($\tilde{T}_\alpha \geq 0.6$) and 0.2 Rydberg ($\tilde{T}_\alpha \geq 0.2$) and where there are no solutions to both the conditions for $\Theta < 0.727$ and $\Theta < 0.67$ respectively. Choosing a lower value of downstream temperature cut-off is seen to have insignificant effect on the upper bound on the snowplow velocity given by 19 . The vertical line at $\tilde{\varepsilon}_{eff} = 0.032$ shows the lower bound given by 19 for $p_0 = 4$ mbar .

There is also a bound on the normalized shock speed: $\tilde{s} \geq 1.6$ . This represents a correction to the snowplow model applied to the assumed PCS configuration: the magnetic pressure needs to be *less* than 2½ times the "wind pressure" in order to be in compliance with conservation laws and the ionization stability condition as formulated above. Combined with the lower bound on the snowplow velocity, this gives for the shock speed the value $1.26 \times 10^5$ m/s, which is in good agreement [5] with experimentally measured velocity range of $1-2 \times 10^5$ m/s.

The existence of both upper and lower bounds on shock velocity can be related to the fact that the downstream temperature and electric field are both increasing functions of shock velocity but the downstream temperature has a *lower bound* related to the need for maintaining adequate thermal ionization and electrical conductivity to limit magnetic field diffusion and ensure a thin current sheath [6] while the electric field has *an upper bound* related to the need for avoiding the ionization wave [16] running ahead of the shock front leading to decoupling between the shock and the piston [5].

This analysis rests on the assumption that the PCS configuration possesses only the azimuthal component of magnetic field and only the component of plasma velocity normal to the shock front. However, a significant mass flow is known [6,7] to exist in the radial direction, parallel to the shock, which is not taken into account in the above discussion. Presence of an axial magnetic field, azimuthal current density and azimuthal plasma velocity [22,23] would also significantly alter its conclusions. These aspects are beyond the scope of this letter.

In summary, this Letter demonstrates that conservation laws of mass, momentum and energy, including contribution from electromagnetic fields, along with the ionization stability condition, require that the ratio of shock velocity to the snow plow velocity be *of the order of unity* (instead of being *equal to unity* as assumed in the snowplow model) and impose both upper and lower bounds on the velocity of the dense plasma focus sheath, which are dependent on material properties (electric breakdown strength and specific ionization and dissociation energy), leading to its observed near-independence from device parameters such as capacitor bank impedance [14] and device geometry [12,14]. This letter thus provides an unambiguous answer to the question raised by Klir and Soto [14] concerning existence of an upper limit on sheath velocity; more work is, however, required to take into account mass flow parallel to the shock [6,7] in arriving at a quantitative understanding of this upper limit.